# Galaxy Zoo:
# Exploring the Motivations of Citizen Science Volunteers

M. Jordan Raddick (Johns Hopkins University), Georgia Bracey, Pamela L. Gay (Southern Illinois University Edwardsville), Chris J. Lintott (Oxford University), Phil Murray (Fingerprint Digital Media), Kevin Schawinski (Einstein Fellow, Yale University), Alexander S. Szalay (Johns Hopkins University), Jan Vandenberg (Johns Hopkins University)

ABSTRACT
The Galaxy Zoo citizen science website invites anyone with an Internet connection to participate in research by classifying galaxies from the Sloan Digital Sky Survey. As of April 2009, more than 200,000 volunteers had made more than 100 million galaxy classifications. In this paper, we present results of a pilot study into the motivations and demographics of Galaxy Zoo volunteers, and define a technique to determine motivations from free responses that can be used in larger multiple-choice surveys with similar populations. Our categories form the basis for a future survey, with the goal of determining the prevalence of each motivation.

## 1. INTRODUCTION

Science educators are increasingly recognizing the key role that participation in authentic scientific practice plays in science education (Michaels, Shouse, and Schweingruber 2008). Participation in scientific research can have many advantages for learners, including a first-hand experience with the scientific process, an increase in knowledge about the research topic (Brossard et al., 2005), and an increase in scientific thinking (Trumbull et al. 2000). However, non-experts often find high barriers to participating in genuine scientific research, such as the need for deep understanding of scientific content and methods, difficulty in accessing data and tools, and inability to make contact with professional scientists who could serve as mentors (Scott, Asoko, and Leach 2007; Sung et al. 2003; Hawkins 1978). Some of these barriers can be lowered when learners work with researchers by participating in carefully-planned citizen science activities.

The practice of “citizen science” involves members of the public (“citizen scientists”) working with professional scientists to complete a research project. The American Association of Variable Star Observers (AAVSO; Williams 2001) and the Audubon Society’s Christmas Bird Count (Root 1988) have both successfully partnered citizen scientists with professional scientists for more than a century, producing research results that could not otherwise have been achieved. More recently, the Cornell Lab of Ornithology has recruited birdwatchers as citizen scientists for published studies of the distribution of bird populations in North America; these studies would have been impossible without participation of citizen scientists (Bonney 2008, Bhattacharjee 2005).

The availability of large scientific datasets through the Internet has allowed citizen science projects to engage volunteers in new ways. In addition to collecting data, citizen scientists can analyze existing data, including data from modern astronomy surveys and missions.

In this paper, we discuss an online citizen science project called Galaxy Zoo. This project asks volunteers – self-styled "Zooites" – to morphologically classify selected images of galaxies from the Sloan Digital Sky Survey. As of April 2009, more than 200,000 people from approximately 170 countries had been involved in making more than 100 million classifications of galaxies. These results are being used in more than 50 research projects that have thus far produced 16 papers accepted or submitted to peer-reviewed journals (see http://www.galaxyzoo.org/story for an up-to-date list of science papers accepted and submitted using data from Galaxy Zoo).

The Galaxy Zoo project offers two interfaces. Galaxy Zoo (GZ1; http://zoo1.galaxyzoo.org) asks users to determine if a galaxy is spiral or elliptical, and to determine the rotation direction of spirals (see figure 1). For a fuller discussion of Galaxy Zoo 1, including a statistical discussion of the accuracy of classifications made by volunteers, see Lintott et al. (2008). A newer project, Galaxy Zoo 2 (GZ2; zoo2.galaxyzoo.org), presents users with more detailed questions about a subset of 250,000 galaxies; GZ2 will be the subject of a future series of papers.

In this paper, we present the history of user involvement in GZ1 (section 2); place our study in the context of other studies that have explored the motivations of volunteers in citizen science projects (section 3); outline our methodology for exploring GZ1 volunteers' motivations; present the categories of motivation we uncovered in the course of this research (section 4); and discuss our confidence in our motivations and future refinements to our study (section 5). The work described in this paper is a pilot for a larger study in which about 10,000 Galaxy Zoo volunteers completed a survey – the motivation categories defined in this paper were used to create the survey questions. The way that this study impacts our ongoing survey research is described briefly in section 5; results from the survey will be described in a future paper.

## 2. HISTORY OF GALAXY ZOO PARTICIPATION

The GZ1 website launched on July 8, 2007. As part of its launch, principal investigator Chris Lintott appeared on BBC's morning radio program, "Today," on July 11th, and the project was reported on the BBC News website (http://news.bbc.co.uk/1/hi/sci/tech/6289474.stm). The project was also announced via his blog, "Chris Lintott's Universe" (www.chrislintott.net).

News of the project was quickly picked up across print and online media, with articles appearing in print in *The Christian Science Monitor* and the *Boston Globe*, and online on *The (London) Times Online*, *badastronomy.com*, and *universetoday.com*, as well as in many other print and online media. As a result of this publicity, news of the project quickly spread -- especially via online media -- and participation increased drastically.

Almost immediately after the BBC radio show coverage aired, tens of thousands of volunteers joined. By the end of the next day (July 12, 2007), nearly 1.5 million classifications had been completed by more than 35,000 volunteer classifiers. For comparison, a graduate student working on a related project was able to classify a sample of 50,000 galaxies in one week, but only by devoting himself entirely to the task (Schawinski et al. 2007). In the nearly 2 years since its original launch, GZ1 has continued to grow, both in number of users and in number of galaxies classified (see figure 2a-b).

Hand in hand with the flood of galaxy classifications came a flood of e-mails. In response to this deluge, the Galaxy Zoo team launched a forum to encourage volunteers to communicate with one another and answer each other's questions (the forum required a separate registration from the main website). The team later launched a blog in which Galaxy Zoo team members (nicknamed "Zookeepers" by volunteers) describe the research they are doing with volunteers' classifications. As of April 11, 2009, the forum had more than 11,000 members, and the blog receives an average of about 25,000 unique visitors per month. One unanticipated consequence of the Galaxy Zoo Forums was the development of novel collaborative research projects by volunteers (Caradamone et al. in press), and the discovery of at least one truly unique object (Lintott et al.2009).

The spectacular growth of GZ1 shows that the site was successful in attracting volunteers to classify galaxies. Additionally, some volunteers go beyond the basic task of classifying galaxies to other thoughtful interactions that may lead to increased understanding of science, such as participating in the forum and blog or participating in collaborative research projects

The success of Galaxy Zoo represents a major opportunity for astronomy research and astronomy education. To take full advantage of this opportunity in future citizen science projects, we must understand what made Galaxy Zoo such a successful project. We must understand what factors motivate volunteers to participate in Galaxy Zoo. By understanding these motivations, we can plan future citizen science projects to appeal to these motivations, potentially maximizing their numbers of participants (and therefore their scientific impact), and increasing public understanding of and participation in science.

## 3. THE CITIZEN SCIENCE LANDSCAPE

As a starting point in trying to understand volunteers' motivations, we review the existing literature concerning volunteer motivation in citizen science. For a more complete literature review of Citizen Science, see Bracey (2009).

Citizen science pre-dates the Internet, and early projects involved participants making observations of the natural world that were often reported through the mail. Today, these observations are often reported through the Internet, but they do not require the Internet (for example, see http://www.history.noaa.gov/legacy/coop.html ; http://www.audubon.org/bird/cbc/history.html ).

The Internet, however, allows citizen scientists to participate in research in completely new ways. For the first time, large numbers of people are being asked to analyze entirely online datasets rather than to create datasets, and to communicate with others entirely online. An important precursor to the host of new online-analysis citizen science projects is SETI@home (Anderson et al. 2002), which launched in 1999. More accurately described as distributed computing, SETI@home asks participants to donate idle time on their home computers to analyze radio telescope data with the hope of discovering signals from extraterrestrial civilizations. In 2006, one of the first entirely online citizen science projects, Stardust@Home (Mendez 2008), invited participants to actively use a web interface to search for dust grains captured in aerogel by a spacecraft sent to study Comet Wild 2. Results from the citizen scientists were used by mission scientists. When a Stardust@Home user discovers a dust grain, he or she is listed as a coauthor on the paper announcing the discovery, and also is given the privilege of naming the dust grain. The Stardust@Home team also writes a News section of their website communicating their results and plans to volunteers (http://stardustathome.ssl.berkeley.edu/news.php).

While citizen science itself has a long history, little research has gone into who makes up the citizen science community and what motivations they have for participating. A few studies appeared in the 1990s, but the majority of the research has only been published in the current decade. Most studies focused on the benefits of citizen science to its participants (both scientists and citizens), and to society. The few studies that also looked at who volunteers and why were conducted with environmental or biology-related projects, and no studies were found that involved completely online projects like Galaxy Zoo. While this pilot study is the first to study the motivations of volunteers in an online astronomy project, an overview of the existing studies will give some context and background to our study of volunteer motivations.

Bradford and Israel (2004) analyzed the types of volunteers involved in the Florida Fish and Wildlife Conservation Commission (FWC) annual survey of Florida beaches for turtle nests. Volunteers participated in a variety of tasks in the sea turtle project, but the most common tasks were to patrol beaches for sea turtle nesting activity, and to mark the nests. The researchers administered a survey containing Likert Scale values for 30 motivation-related questions that was returned by 382 volunteers. Volunteers in the FWC survey tended to be older, well-educated, white females. Most people volunteered for more than one reason, but the most popular reason by far was to help and protect sea turtles.

King and Lynch (1998) studied volunteers working at the Ohio Chapter of the Nature Conservancy who participated in a variety of environmental management tasks, including helping with studies of nesting birds. The study used a 14-question survey; the first question provided a list of 12 motivations and asked volunteers to check those that applied to them. Another question asked volunteers to report which single motivation applied most to them. A total of 86 people responded to the survey. The researchers found that these volunteers were mostly middle-aged, well-educated, fairly affluent white males. About three-quarters of this group worked full-time, and 40% had completed a graduate degree. Most respondents checked multiple motivations for volunteering, but when choosing the single motivation that was most important to them, 63% said they wanted "to do something for nature." Other motivations included "to explore career options," "to help create a better society." and "to allow the organization to provide more goods/services for less money."

It is important to know whether these findings can be generalized to participants in other types of citizen science projects; for example projects concerning other areas of science or involving different types of technology, and whether they can be generalized to the new generation of online data analysis citizen science projects like Stardust@home and Galaxy Zoo. A goal of our study is to determine the demographic characteristics and motivations of participants in the Galaxy Zoo citizen science project. Comparison of our results with results of these studies will show whether the same motivations hold in an online astronomy project.

## 4. METHODOLOGY FOR DETERMINING MOTIVATIONS

To determine a list of Galaxy Zoo 1 volunteer motivations, we looked for trends in free-form responses gathered during two different solicitations of information. During our first phase, volunteers were asked to respond publicly in the Galaxy Zoo forum (see section 4.1), and during the second phase, selected volunteers responded to interviews (see section 4.2).

### 4.1 Forum Survey

As a preliminary mechanism to learn about volunteers' motivations from the volunteers themselves, the team created a new topic on the site's forum titled "What Makes Galaxy Zoo Interesting?" The introductory message to this forum topic, posted by science team member Kate Land, read:

> *We have been overwhelmed by the awesome response we have had to this project. We feel like we've really captured people's imaginations with Galaxy Zoo, and we'd like to know why.*

*Please leave us feedback here about why you are taking part. We hope that what we learn from you can be helpful to future projects like ours.*

As of January 22, 2009, the day we downloaded our final dataset of forum posts for analysis, this forum message had generated 826 responses. The richness of the forum responses suggested that we should begin a more rigorous study of volunteers' motivations to participate in Galaxy Zoo. This more rigorous study involved interviewing Galaxy Zoo volunteers, and is described below in sections 4.2-4.5. We read the initial forum responses in December 2007 before creating the interview protocol, but a rigorous analysis of the 826 forum posts was conducted only after interview transcripts were analyzed.

4.2 Selecting Interview Subjects

After securing permission from the Institutional Review Board at Johns Hopkins University, we conducted half-hour interviews with Galaxy Zoo volunteers – volunteers who participated in the citizen science activity (classifying galaxies), but not necessarily in the forum. The goal of these interviews was to generate from the interviewees' free responses a set of motivation categories from an initial sample of Galaxy Zoo participants. We interviewed 22 people out of a total of about 160,000 Galaxy Zoo volunteers.

To obtain a pool of interview subjects, we sent out an e-mail solicitation to volunteers who had chosen to register their e-mail with the site, and who had not opted out of receiving Galaxy Zoo mailing list e-mails. E-mail solicitations were sent out in four batches of approximately 300-400 e-mails each. We had originally hoped to obtain enough response for about 20 interviews from the first batch of solicitations, but getting this amount turned out to require four separate attempts. This solicitation was consistent with the Galaxy Zoo privacy policy, which states that volunteers may be contacted for the purpose of research into the site's operation. The total number of volunteers who received solicitations for interviews was 1,336.

One potential area of concern was the question of whether we might contaminate further studies into volunteers' motivation by soliciting some volunteers for interviews in this study, then asking those same volunteers about their motivations in a future survey (section 5). However, the large size of the Galaxy Zoo volunteer population mitigates this concern. When the first e-mail solicitation was sent on April 1, 2008, the number of volunteers had risen to 137,228; by the time the last solicitation was sent on August 24, 2008, it had grown to 161,961. The 1,336 volunteers who received email solicitations to participate in the interviews described here are less than 1% of the overall Galaxy Zoo volunteer population, meaning that less than 1% of our population will potentially be asked about their motivations twice. We believe that this will not pose a significant threat to the internal validity of either this study or the survey study that will follow.

As an inducement for participating in an interview, we promised to send five interview subjects, selected at random, a copy of an astronomy book signed by the authors, one of whom is a Galaxy Zoo research team member. E-mails to use as contacts for this offer were kept separately from all information used in this study, with no links between subjects' interview transcripts and their E-mails.

When volunteers responded to the e-mailed interview request, a mutually convenient time was set for an interview. Interviews were scheduled to take half an hour. As mentioned above, interviews were conducted with 22 subjects. (An additional 21 people offered to participate, but either did not return e-mails or could not find a mutually convenient time for both interviewer and interviewee to talk.) Table 1 shows demographic characteristics for all participants included in the study.

INSERT HERE Table 1. Volunteers interviewed for this study. The table shows the interview code used by the authors, the type of interview (IM or Phone), and the interviewee's age, gender, country of residence, and occupation (occupations were adapted from volunteer self-reports except where in quotes).

| Code | Type | Age | Gender | Country | Occupation |
|---|---|---|---|---|---|
| A | IM | 58 | Male | UK | Technology / IT project manager /web designer |
| B | IM | 57 | Male | UK | IT, as CIO of a large banking group |
| C | IM | 25 | Male | US | Build and configure computers |
| D | IM | 49 | Male | Denmark | "I work for a company doing engineering software" |
| E | IM | 47 | Male | US | Sheet metal fabricator |
| F | IM | 61 | Female | UK | Surgical Nurse |
| G | IM | 25 | Male | Netherlands | PhD student in astronomy |
| H | IM | 24 | Male | France (UK citizen) | Environmental advisor for oil company |
| I | IM | 24 | Male | UK | Physics teacher |
| J | IM | 23 | Male | US | Investment banking |
| K | IM | 25 | Female | Argentina | Freelance web developer |
| L | IM | 51 | Male | Belgium | Chemist |
| M | Phone | 42 | Female | US | [Declined to report] |
| N | IM | 30 | Male | Japan | [Declined to report] |
| O | Phone | 45 | Male | US | Scientific illustrator |
| P | Phone | 68 | Male | UK | Retired |
| Q | IM | 38 | Male | US | Software quality assurance |
| R | IM | 36 | Female | US | "was pre-med, … later…a math major" |
| S | Phone | 79 | Male | UK | Automotive engineering business |
| T | IM | 60 | Male | US | Judge (state court) |
| U | Phone | 56 | Female | UK | Draftsperson in theater |
| V | IM | 22 | Male | France | Engineering student |

4.3 Interview Protocols

Our initial interview protocol is shown in Appendix A. We asked a series of questions about subjects' demographics, their impressions of the Galaxy Zoo website, their motivations for participating, and their experiences with and definition of science. (The question about definition of science was taken from the *Views on Science-Technology-Society* instrument [Aikenhead and Ryan 1992]).

As an online project, Galaxy Zoo is likely to attract a technically literate audience. With this audience, we thought that interviews using a technical medium like instant messaging might be appropriate, and some research shows that instant messaging can be an effective medium for conducting interviews in human-computer interaction research (Voida et al. 2004). Some subjects were not comfortable using instant messaging software, so phone interviews were conducted with these subjects. The same researcher (Raddick) conducted all interviews, both instant messaging and phone. All phone interviews were recorded and later transcribed. Questions were asked in the same order in the first seven interviews, but in some interviews, questions were skipped to keep the total interview within the half-hour timeframe the volunteer had committed.

After conducting these seven interviews, with codes A through G, we realized that some questions about participants' past experiences with and attitudes toward science -- questions we considered important -- were not being asked in every interview. We therefore changed the order in which questions were asked to bring these questions closer to the beginning of the interview; the revised order is shown in Appendix B. The protocol shown in Appendix B was used to conduct interviews with codes H through V. All the results reported here come from analysis of questions that were asked on both versions of the survey protocol, and therefore asked of all subjects. Results of our analysis of questions concerning their experience with and definition of science are left for future work.

### 4.4 Analysis of Responses – Creating the Categories

To ensure that our motivation categories accurately reflected what people said in the interviews, we used a grounded theory approach in which theoretical models – in our case, categories of motivation – emerge from the data (Strauss and Corbin 1990). We began by having two independent raters (Raddick and Bracey) read through the first 12 interviews (those with codes A through L; we chose these twelve because they were the interviews we had already conducted at the time we began this analysis). Each rater read through each interview and identified statements of motivation in the interview transcripts. A statement of motivation is a phrase in the transcript that provides insight into a reason why the subject participated in Galaxy Zoo. Phrases were considered independently; for example, the hypothetical statement, "As a person who loves astronomy, I enjoy looking at pictures of the night sky" would be considered two separate statements of motivation: "A person who loves astronomy" and "I enjoy looking at pictures of the night sky." Rater #1 (Raddick) identified 79 statements of motivation; rater #2 (Bracey) identified 80 statements of motivation.

Each rater independently wrote the identified statements of motivation onto separate index cards. In many cases, one person used multiple statements to refer to what was later grouped into a single motivation by the raters. In this way, an interview subject might have ten statements of motivation that refer to three motivations. Once the motivational statements were recorded, the raters then shuffled the index cards and sorted them into categories by identifying common themes within the statements of motivation (Tables 2a and 2b). In addition, a third rater (Gay) read through the transcripts of the first ten interviews (those with codes A through J) and identified groups of statements of motivation (e.g. all the phrases related to one motivation) in each interview. The rater then sorted these groups of statements into motivation categories (Table 2c).

INSERT HERE: Table 2. Initial categorization scheme for motivations developed by each rater. The table gives the rater's name for the motivation category and a typical statement of motivation from that category. a) Schema of rater #1, b) Schema of rater #2 , and c) Schema of rater #3.

**Table a.**

| Motivation Category | Typical Statement of Motivation from category |
|---|---|
| Astronomy | I have an interest in astronomy. |
| Fun | I had a lot of fun categorizing the galaxies :-) |
| Helping | I thought that if I could help I should… |
| Images | I liked the pictures as well. |
| Imagination | It stretches the imagination |
| Interesting Project | It's a very interesting study… |
| Scale of the Universe | Well he [the subject's son] has gotten a better idea of the vast size of the universe |
| Science | Always had an interest in science… |
| Teaching | I'm a physics teacher and a member of the PTNC [Physics Teacher News & Comments] group. |
| Miscellaneous | …the slight probability that I may point out the one object that will completely shock our current understanding about the universe |

**Table b.**

| Motivation Category | Typical Example |
|---|---|
| Desire to learn | family project for me and my son… who has a desire to learn about astronomy |
| Fun | a lot of fun |
| Helping | happy to help |
| Interested in project itself | thought it was an interesting idea to utilize the Internet to bring together tons of people to help classify these galaxies that would otherwise take up a lot of time |
| Interested in science/astronomy | Astronomy has always interested me |
| Involvement in "real science" | participate in a real scientific project |
| Liked the pictures | I like the pictures as well |
| Sense of wonder | partly the romantic side of so many possibilities of different environments, life, planets, stars |
| Not used | I think that the user interface for classification is pretty well done. It is simple and straightforward. |

**Table c.**

| Motivation Category | Typical Example |
|---|---|
| Scale of Universe | So many galaxies, so many in a single image, and the beauty of them |
| Images | I enjoyed having a flick through the photos |
| Astronomy | I enjoy astronomy quite a bit. … So obviously, seeing a project focused on galaxy classification garnered my interest. |
| Science | First, I enjoy studying the physics in astronomy |
| Volunteerism | I thought that if I could help I should |
| Fun | For the fun of it |
| Teaching | I'm a physics teacher … It looked like it would be a good way to communicate the vastness of the universe. |
| Galaxy Zoo | I was interested in seeing how 'human computing' would be used in the project. |

In summary, rater #1 identified nine separate motivations from the statements of motivation; rater #2 identified eight separate motivations; and rater #3 identified eight separate motivations. The next step was for the three raters to discuss all these motivations and determine which motivations were stating the same concepts in different words. We concluded that the following identified categories could be combined or split:

- Rater #2's "Interested in Science / Astronomy" category was split to correspond to rater #1 and rater #3's separate "Astronomy" and "Science" categories.
- Rater #1's "Imagination" category and rater #2's "Sense of wonder" category were considered aspects of the same motivation expressed by reviewer #3's "Scale of the Universe" category; this motivation was given the name "Vastness"
- "Helping" (raters #1 and #2) and "Volunteerism" (#3) were considered the same category, referred to as "Helping"
- "Images" (raters #1 and #3) and "Liked the pictures" (#2) were considered the same motivation, referred to as "Beauty"
- Raters #1 and #3 agreed that rater #2's "Involvement in 'real science'" category was a separate from "Helping"; this motivation was referred to as "Contribute"

As a result of these discussions, we settled on a list of twelve motivations, and a typical statement of motivation describing each; these statements were used to describe each motivation on our survey. These twelve final motivations are shown in Table 3; the table shows the motivation category names used by the research team and the one-sentence descriptions used in the survey described in section 5.

INSERT HERE Table 3. Final motivation categories that arose during the interviews, selected after discussion among the three raters. The table shows the motivation category name as used by the research team and the one-sentence description of the category used in the survey.

| **Motivation Category** | **Description (used in survey instrument)** |
|---|---|
| Contribute | I am excited to contribute to original scientific research. |
| Learning | I find the site and forums helpful in learning about astronomy. |
| Discovery | I can look at galaxies that few people have seen before. |
| Community | I can meet other people with similar interests. |
| Teaching | I find Galaxy Zoo to be a useful resource for teaching other people. |
| Beauty | I enjoy looking at the beautiful galaxy images. |
| Fun | I had a lot of fun categorizing the galaxies. |
| Vastness | I am amazed by the vast scale of the universe. |
| Helping | I am happy to help. |
| Zoo | I am interested in the Galaxy Zoo project. |
| Astronomy | I am interested in astronomy. |
| Science | I am interested in science. |

Once we settled on this list of twelve motivation categories, the next step was for each of the three raters to code the entire set of interview transcripts, as well as ten additional interviews that were conducted later (those with codes M through V), using these motivation categories. This step allows us to calculate the inter-rater reliability of the categories (section 4.5).

To ensure that we were not missing any major motivation categories, we used the same motivation categories to classify all motivations identified within the forum posts (see section 4.1), with an eye toward identifying any missing categories. Within the 826 forum responses, 215 responses contained statements of motivation, and two were thrown out because responders indicated they were under the age of 18. (While we have no way of knowing the ages of other forum responders, we proceeded under the assumption that since the forum posts are already in the public domain, no harm would be done to forum participants who did not provide demographic information and whose motivations were kept separate from their forum username within the spreadsheets we used in our analysis.)

After analyzing all forum responses, we did not identify any potentially new motivations that the raters could not independently fit within an existing category. Without being able to crosscheck our list of motivations against a large sample of forum posts, it would have been difficult to conclude that we had a complete list of motivations. Because we did not find any additional motivations in the larger sample of forum posts, we conclude that we do have a nearly complete set of volunteer motivations. In addition, since forum users and e-mail respondents would be affected by different biases, the fact that these different groups share the same motivations adds weight to the generalizability of our samples. Detailed inter-rater reliability and response rates are discussed below.

Both the forum and interview data are limited in ways that make the determination of the frequency of motivations difficult. In the forums, posters often referenced motivations from others' prior posts in what they said, and the motivation distribution of people who use the forums is not necessarily the same as the motivation distribution of the much larger group of volunteers who only classify galaxies. In addition, although interviews were sought from a randomly chosen set of volunteers who would ideally represent an unbiased sample, we recognize that the sample will by its nature only reflect the motivations of people who agreed to be interviewed. In this case, of the 1,336 people we invited to participate, only 22 were interviewed, a response rate of only 1.4 percent. However, the goal of this study was to determine a complete set of motivation categories to be used in further research; accomplishing this goal does not require us to make any statements about the frequency with which motivations appear in our population.

## 4.5 Inter-rater reliability

As a check on inter-rater reliability, we went through the interview transcripts and forum posts, and categorized just the first motivation that the volunteer mentioned. Since both these datasets (all interviews and forum posts) required the raters to examine free-form text-based data (either interview transcripts or forum posts), we were able to combine both datasets for the purposes of measuring inter-rater reliability.

Overall, we found that for the complete dataset, all three of the raters matched for 644 (76%) classifications, two out of three matched for 175 (21%) of the classifications, and in only 29 cases (3%) was there no agreement. However, these values may not be the best indication of inter-rater reliability. This is because the full data set includes 609 forum posts for which 2 or more raters indicated that no statement of motivation was present.

After excluding these items and considering only the remaining 237 responses, we found all three raters matched 98 times (41%) and two out of three matched for 110 (46%) of the items, and no agreement was found in the same 29 (12%) items. This corresponds to a pairwise percent agreement of 56% between raters #1 and #3, 56% between raters #1 and #2, and 58% between raters #2 and #3, as well as a Fleiss Kappa value of 0.474. For the purposes of being able to identify what motivations are present consistently, but not for identifying the frequency of motivations to high accuracy, we consider our results reliable.

# 5. DISCUSSION

It is important to emphasize that most users do not have a single motivation. After eliminating forum posts that were not relevant (for instance "Thanks for responding" messages), we found the majority of the respondents mentioned more than one motivation within either their interview or forum posting, with the typical respondent listing two motivations (102 forum and seven interview respondents, or 46% of the combined sample of interviews and forum posts). Only 89 (37%) of the 237 posts and no interviews included statements of motivation that all raters could fit into only one category. This is consistent with the previous studies of citizen science volunteer motivation discussed in section 3, which found that most volunteers reported more than one motivation.

We examined the motivations in the order they are stated to check inter-rater reliability (section 4.5). There will not necessarily be a direct correlation between a respondent's primary motivation -- the motivation that they consider most important -- and what they say first. While one might reasonably guess that the order that motivations are listed may correlate to importance of the motivations to the responders, we have no data to check this hypothesis.

We next looked at the frequencies with which these motivations appeared in our sample (Table 4). As discussed in section 4.4, these reported frequencies do not indicate the prevalence of these motivations in the larger volunteer population, but the fact that the motivations appear with reasonable frequency in this sample does give us confidence that these motivations are truly present in the larger population.

INSERT HERE: Table 4: Frequency of motivations identified in respondents' interviews or forum posts. Only motivations found by two or more of the three raters are counted. The 631 responses determined "non-pertinent / containing no motivations" by two or more raters were removed from the sample prior to calculating percentages. Percentages are based on the number of times among the 207 remaining responses containing a motivation. This includes both the forum posts and the 22 interviews. These frequencies do not indicate the frequencies that these motivations appear in the larger population, but they do give us confidence that these motivations are present in the larger population.

| **Motivation** | **Interviews** | | | | **Forums** | | | | **All** | | | |
|---|---|---|---|---|---|---|---|---|---|---|---|---|
| | **Initial responses** | | **All responses** | | **Initial responses** | | **All responses** | | **Initial responses** | | **All responses** | |
| | **freq** | **%** | **Freq** | **%** | **freq** | **%** | **freq** | **%** | **freq** | **%** | **freq** | **%** |
| Astronomy | 7 | 35% | 11 | 17% | 86 | 46% | 99 | 46% | 93 | 39% | 110 | 46% |
| Beauty | 1 | 5% | 4 | 6% | 9 | 5% | 35 | 16% | 10 | 4% | 39 | 16% |
| Community | 0 | 0% | 1 | 2% | 3 | 2% | 14 | 6% | 3 | 1% | 14 | 6% |
| Contribute | 1 | 5% | 6 | 9% | 29 | 15% | 46 | 22% | 30 | 13% | 52 | 22% |
| Discovery | 1 | 5% | 3 | 5% | 4 | 2% | 17 | 8% | 5 | 2% | 20 | 8% |
| Fun | 3 | 15% | 5 | 8% | 14 | 7% | 21 | 11% | 17 | 7% | 26 | 11% |
| Help | 0 | 0% | 6 | 9% | 4 | 2% | 10 | 7% | 4 | 2% | 16 | 7% |
| Learning | 1 | 5% | 3 | 5% | 5 | 3% | 21 | 10% | 6 | 3% | 24 | 10% |
| Science | 0 | 0% | 4 | 6% | 2 | 1% | 5 | 4% | 2 | 1% | 9 | 4% |
| Teaching | 1 | 5% | 1 | 2% | 2 | 1% | 3 | 2% | 3 | 1% | 4 | 2% |
| Vastness | 1 | 5% | 10 | 15% | 25 | 13% | 47 | 24% | 26 | 11% | 57 | 24% |
| Zoo | 4 | 20% | 11 | 17% | 5 | 3% | 8 | 8% | 9 | 4% | 19 | 8% |
| Non-pertinent | | | | | 609 | N/A | | | 609 | N/A | | |

In looking at the distribution of initial responses where two or more raters agreed as shown in Table 4, we found that only three responses (after removing "non-pertient" / "no motivation" from the sample) appeared at a 10% or greater level among the list of first-mentioned motivations: Astronomy (39%), Contribute (13%), and Vastness (11%).

Expanding the sample to include all motivations listed by each person rather than just the first, and again only considering cases where two or more raters agreed, we find that these motivations continue to dominate, but more motivations are found in common across more than 10% of the responders.

The primary goal of this investigation was to identify a list of motivation categories driving participants in the GZ1 project to classify galaxies. We have used these categories to construct a survey instrument that uses multiple-choice and Likert scale responses to assess the frequency of each motivation category in our population. We have administered this instrument as an online survey and are now analyzing the responses. Results of this survey research will be published in a future paper.

While we were able to match all the motivational statements in both the interviews and the forum free responses against the twelve motivation categories, we recognize that our sample has certain biases that may indicate that there are motivations we could not have sampled. Our only comparison dataset was posts to the Galaxy Zoo forum; however, only about 5% of all Zooites have a forum account, and not all of those individuals are active within the forums. There may also be unmeasured population differences between those users that self-select to socially and intellectually interact in the forums and users who only classify galaxies. For example, the motivation to participate in GZ1 as a mechanism to make friends is much less likely to be present in people who do not socially engage through the Galaxy Zoo forums. Recognizing that the comparison data set is not an unbiased sample, the survey that we are now conducting will allow users the opportunity to offer a free response motivation in addition to asking them to select the relative importance of each of the previously identified motivations.

## 6. CONCLUSIONS

Our method of interviewing participants, identifying motivational statements, and categorizing those statements into a set of overarching motivations produced a list of motivations that was applicable across both an extended interview set and forum posts. We conclude that it is possible to explore the motivations of a large population of citizen scientists through a very small sampling of individuals using interviews, although we recommend using larger interview sets in future studies.

Through interviews of 12 randomly selected GZ1 participants, we were able to identify approximately 80 "statements of motivation" describing reasons why subjects chose to participate in Galaxy Zoo. Three raters classified these statements of motivation, and consensus among the raters led to 12 categories of motivation. Each rater then classified each interview transcript into all the motivation categories that applied to it, noting which motivation category the subject mentioned first. We used these rater classifications to measure the interrater reliability of our categories.

After our initial analysis, we searched for additional motivations in 10 additional interviews and in 826 forum posts, and we did not identify any additional motivations. We feel that being able to cross-check our interview based motivations list against the forum posts provided adequate levels of certainty that we have identified a list of motivations sufficiently representative of the majority of Galaxy Zoo volunteers that we can proceed with a broader future study with a larger sample. We are now conducting this study by surveying a large sample of volunteers. Results from this larger study will offer insight into why people choose to participate in citizen science activities, which will be useful in designing future citizen science projects.

## ACKNOWLEDGMENTS

This publication has been made possible by the participation of more than 100,000 volunteers in the Galaxy Zoo project.

We would like to acknowledge the help and support that Tom Foster provided during the earliest stages of this work, and the input of Karen Carney. Xiaoquan Zhang, Feng Zhu, and Aaron Price provided valuable input on methodologies. This project would not have been possible without the technical support of Johns Hopkins University, who hosted and advised GZ1, and that of Phil Murray (Fingerprint Digital Media) and Edward Edmondson (University of Portsmouth), who have managed and hosted the Galaxy Zoo forums, and Alice Sheppard and the other forum moderators.

Support for the work of KS was provided by NASA through Einstein Postdoctoral Fellowship grant number PF9-00069 issued by the Chandra X-ray Observatory Center, which is operated by the Smithsonian Astrophysical Observatory for and on behalf of NASA under contract NAS8-03060. KS gratefully acknowledges support from Yale University.

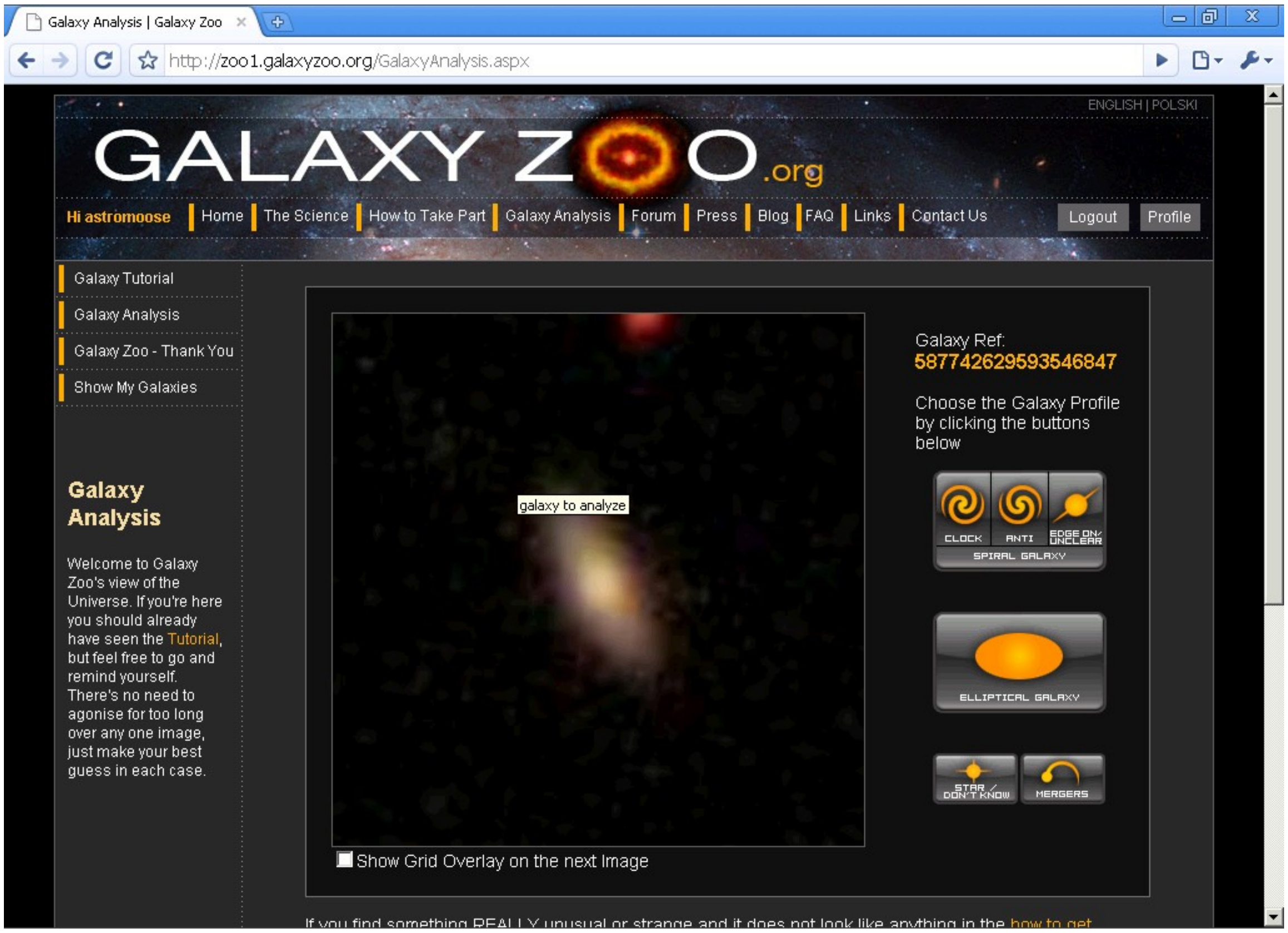

**Figure 1**: A screenshot of the Galaxy Zoo interface. Volunteers look at the galaxy in the center of the screen and determine its shape. They then click one of the six buttons in the right side of the screen to report their classification. Their report is written into a database and compared with the findings of other volunteers to create a database of galaxy morphologies.

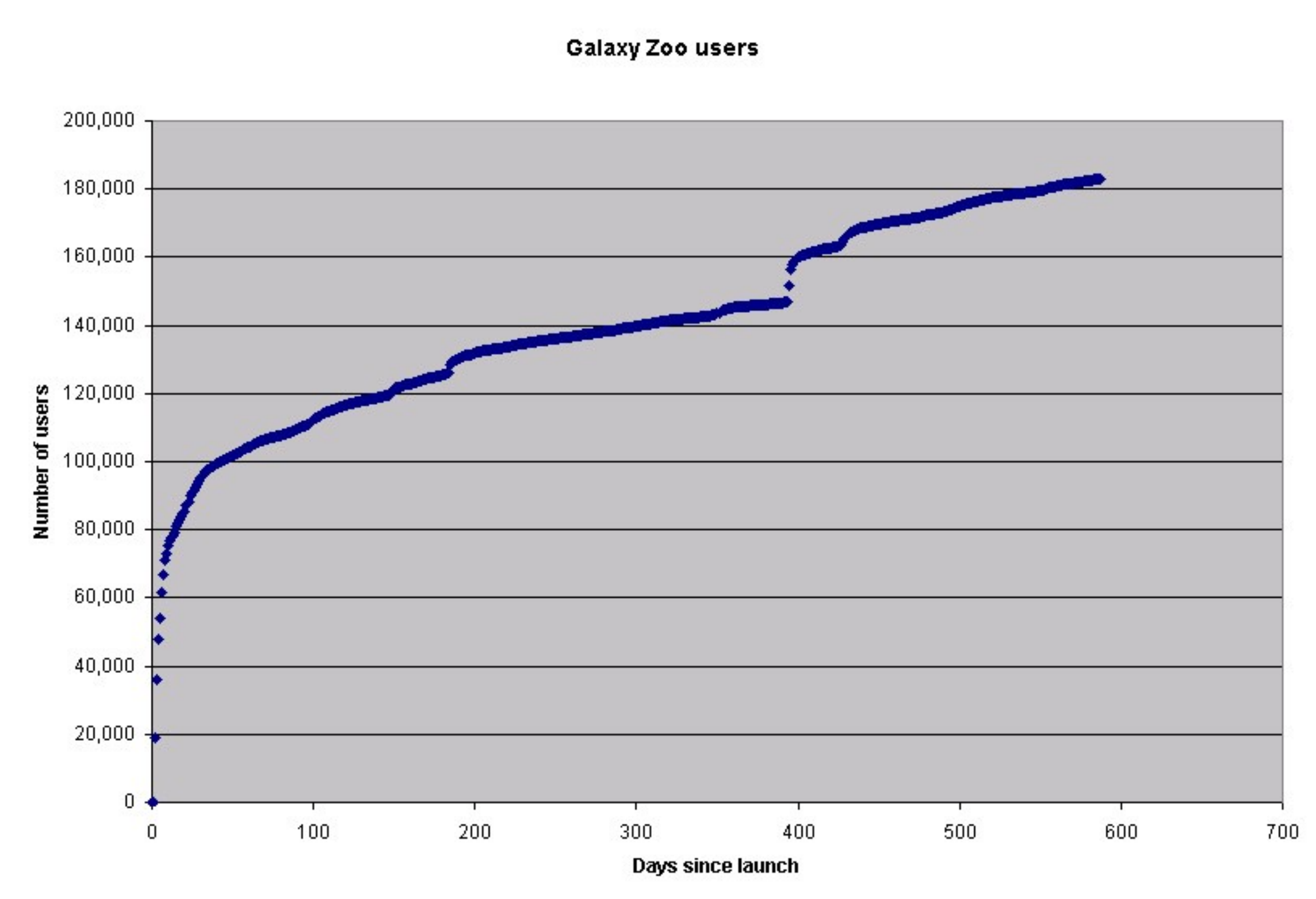

**Figure 2a:** Number of days since the launch of GZ1 versus cumulative number of registered users. The large jump in users on day 394 corresponds to press coverage of the discovery of a new object called "Hanny's Voorwerp" (Lintott et al. 2009).

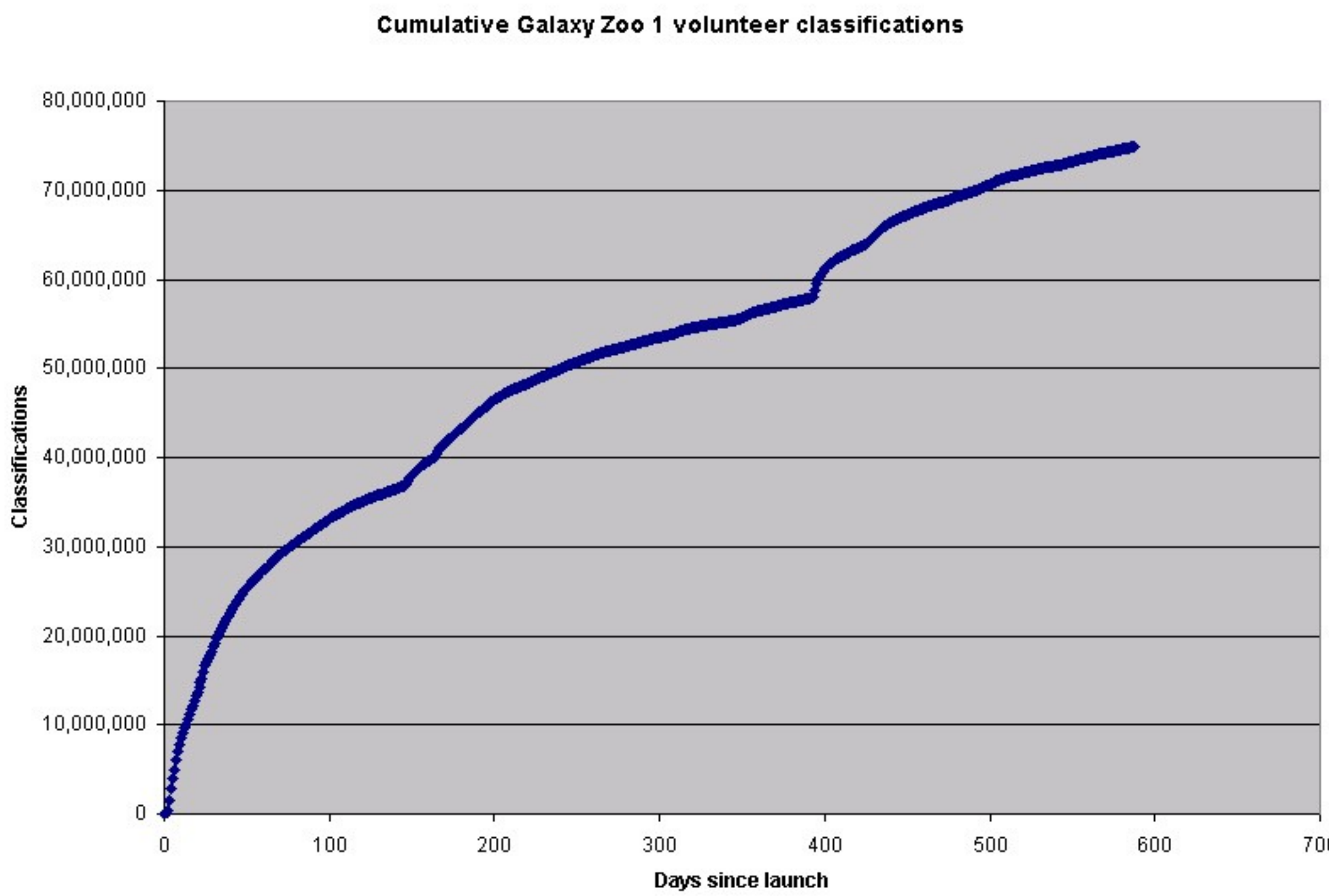

**Figure 2b:** Number of days since the launch of GZ1 versus cumulative number of volunteer classifications. The same large jump on day 394 also appears in the number of classifications.

## APPENDIX A: INITIAL INTERVIEW PROTOCOL FOR INTERVIEWS #1-12

Hello! My name is Jordan, and I'll be asking you a few questions through [Skype/IM/Messenger] today. Thank you for your interest in Galaxy Zoo, and for agreeing to participate in this interview. This interview contains 24 questions, and should take about 30 minutes to complete.

Your responses will help us improve the site to better meet your needs. Before we start, have you read the Information Sheet on the website?

[if no:] Please take a moment to read the sheet. You can find it at http://www.galaxyzooblog.org/wp-content/uploads/2008/04/interview.swf .

Let me explain what I will be asking you. There are six parts to this interview.
First, I will ask some basic demographic questions.
Second, I will ask about why you have chosen to participate in Galaxy Zoo.
Third, I will ask your opinion on some features of the site.
Fourth, I will ask for your responses to some "thought experiments" – if we changed the site in a few different ways, what would you think?
Fifth, I will ask about your experiences with science.
Sixth, I will ask for your definition of science.
As you answer, I may sometimes ask follow-up questions.

Do you have any questions about what we'll be talking about today?

If you have any questions as we go, please ask – I'm happy to answer! If you would prefer not to answer a question for any reason, just say "prefer not to answer," and I will move on to the next question. You don't have to give a reason. If you would like to end the interview at any time, please let me know. Do you have any questions about this?

Great, let's get started! First, let me ask you for some basic data about yourself (5 questions).

### Demographics

1. What is your age? (For legal reasons, this survey is limited to participants over the age of 18.)
    a. [if under 18:] Unfortunately, for legal reasons, we'll have to stop the official interview. Still, thank you for your time! If you have any questions, or you would like a copy of the published research that results from this interview, contact the study coordinator, Jordan Raddick, at +1 410 516 8889.
2. What is your sex (gender)?
3. In what country do you reside?

4. What is your…
    a. [U.S.] ZIP code?
    b. [U.K.] the first part of your postcode?
5. What is your Galaxy Zoo username? (optional – see the information sheet at http://www.galaxyzooblog.org/wp-content/uploads/2008/04/interview.swf about what we will do with this information)

Thanks! Next, let me ask about what you have gotten out of Galaxy Zoo; why you chose to volunteer your time to classify galaxies (which we very much appreciate!). (3 questions)

**Your Motivations**

1. What were your reasons for joining Galaxy Zoo?
    a. [if they give only one reason:] Were there any other reasons?
    b. [if they give several reasons in question #1:] You've named several reasons for participating in Galaxy Zoo. Which one do you consider most important? Why?
2. Are you still classifying galaxies on Galaxy Zoo today?
    a. [yes:] What are your reasons for continuing to classify galaxies today?
    b. [no:] What reasons did you have for stopping?
3. Has your participation in Galaxy Zoo helped you in any "real life" settings, such as at your job?
    a. [if yes] How?

Thank you! Now, let me ask for your opinion on some parts of the website. We are interested in your honest assessments – I didn't write the site, so don't worry about hurting my feelings! (4 questions)

**Evaluating the Site**

1. Where did you first find out about Galaxy Zoo?
2. How difficult do you find classifying galaxies? Is there anything that would make it easier?
3. What comments do you have on the Galaxy Zoo interface?
4. From where do you most often use Galaxy Zoo?
5. What suggestions do you have for future versions of the site?

Thanks! Next, I'll do some "thought experiments." Imagine that Galaxy Zoo were a little different in the ways I'll describe. How would that change your opinion of the project, or how much you would want to volunteer?

**"Thought Experiments"** [note: these will be asked in random order]

1. Astronomers are now analyzing the classifications from Galaxy Zoo users, including you. They will almost certainly discover many interesting things about the universe. But, if you were told that for some reason your classifications were excluded from our study sample, how would that change how you participate in the project? (By the way, this is just hypothetical – all classifications are included in our sample.)
2. Galaxy Zoo depends on people like you choosing to offer their time to classify galaxies. But, if you were told by someone else, such as a boss or a teacher, to participate in Galaxy Zoo, how would that change how you participate in the project?
3. Galaxy Zoo includes about a million galaxies, most of which have been looked at by only a few people. If Galaxy Zoo instead included galaxies that had been extensively studied before, how would that change how you participate in the project?
4. If Galaxy Zoo did not include a forum, so that you could not interact with other users, how would that change how you participate in the project?
5. Some people on the forum have joked that the site is "addictive" – that they can't help doing "just one more galaxy." If we changed the interface so that you had to request the next galaxy (rather than getting the next one immediately), how would that change how you participate in the project?
6. As you know, we are classifying galaxies that have never been classified before, and you are helping us. But if the site gave you a "right" answer after you finished classifying a galaxy, how would that change how you participate in the project?

Thanks – that was very helpful! Next, let me ask a few questions about your experiences with science, past and present (4 questions).

**Your experiences with science**

1. Does your job or course of study involve science, maths, engineering or technology? How?
2. Have you ever studied science? To what level?
3. Are there any science-related activities that you enjoy (for example, stargazing or reading popular science books)? Which ones?
4. Do you participate, or have you participated, in other large online collaboration projects (for example, SETI@home, Stardust@home, Wikipedia, etc.)?
    a. Which ones?

Thank you! Now, the last question is about your definition of the term "science." We are interested in your first impressions, so don't think too long about this – just say the first thing that comes to your mind.

**Your definition of science**

1. How would you define the term "science"?

We've reached the end of the interview! Thank you so much for your help.

Here's what happens next. I will now save this transcript, but I will delete your username and refer to you by a code number. Your code number is XXX. If you'd like to withdraw from the study at any time, contact the study coordinator, Jordan Raddick (raddick@jhu.edu), and give him your code number. If you have any questions, or you would like a copy of the published research that results from this interview, contact Jordan.

You'll be entered into a prize drawing, as explained in my E-mail to you. We'll contact you by E-mail if you have won (but your E-mail will be kept separate from this transcript).

## APPENDIX B: FINAL INTERVIEW PROTOCOL FOR INTERVIEWS #20-47

### Interview Protocol for Galaxy Zoo study

Hello! My name is Jordan and I'll be asking you a few questions through [Skype/IM/Messenger] today. Thank you for your interest in Galaxy Zoo, and for agreeing to participate in this interview. This interview contains 24 questions, and should take about 30 minutes to complete.

Your responses will help us improve the site to better meet your needs. Before we start, have you read the Information Sheet on the website?

[if no:] Please take a moment to read the sheet. You can find it at http://www.galaxyzooblog.org/wp-content/uploads/2008/04/interview.swf .

If you have any questions as we go, please ask – I'm happy to answer! If you would prefer not to answer a question for any reason, just say "prefer not to answer," and I will move on to the next question. You don't have to give a reason. If you would like to end the interview at any time, please let me know. Do you have any questions about this?

Great, let's get started! First, let me ask you for some basic data about yourself (5 questions).

**Demographics**

1. What is your age? (For legal reasons, this survey is limited to participants over the age of 18.)

a. [if under 18:] Unfortunately, for legal reasons, we'll have to stop the official interview. Still, thank you for your time! If you have any questions, or you would like a copy of the published research that results from this interview, contact the study coordinator, Jordan Raddick, at +1 410 516 8889.
2. What is your sex (gender)?
3. In what country do you reside?
4. What is…
    a. [U.S.] your ZIP code?
    b. [U.K.] the first part of your postcode?
5. What is your Galaxy Zoo username? (optional – see the information sheet at http://www.galaxyzooblog.org/wp-content/uploads/2008/04/interview.swf about what we will do with this information)

Thanks! Next, let me ask about what you have gotten out of Galaxy Zoo; why you chose to volunteer your time to classify galaxies (which we very much appreciate!). (3 questions)

**Your Motivations**

1. Where did you first find out about Galaxy Zoo?
2. What were your reasons for joining Galaxy Zoo?
    a. [if they give only one reason:] Were there any other reasons?
    b. [if they give several reasons in question #1:] You've named several reasons for participating in Galaxy Zoo. Which one do you consider most important? Why?
3. Are you still classifying galaxies on Galaxy Zoo today?
    a. [yes:] What are your reasons for continuing to classify galaxies today?
    b. [no:] What reasons did you have for stopping?

Thanks – that was very helpful! Next, let me ask a few questions about your experiences with science, past and present (4 questions).

**Your experiences with science**

1. Does your job or course of study involve science, maths, engineering or technology? How?
2. Have you ever studied science? To what level?
3. Are there any science-related activities that you enjoy (for example, stargazing or reading popular science books)? Which ones?
4. Do you participate, or have you participated, in other large online collaboration projects (for example, SETI@home, Stardust@home, Wikipedia, etc.)?
    a. Which ones?

Thank you! The next question is about your definition of the term "science." We are interested in your first impressions, so don't think too long about this – just say the first thing that comes to your mind.

**Your definition of science**

1. How would you define the term "science"?

Thanks! Next, I'll do some "thought experiments." Imagine that Galaxy Zoo were a little different in the ways I'll describe. How would that change your opinion of the project, or how much you would want to volunteer?

**"Thought Experiments"** [note: these will be asked in random order]

1. Astronomers are now analyzing the classifications from Galaxy Zoo users, including you. They will almost certainly discover many interesting things about the universe. But, if you were told that for some reason your classifications were excluded from our study sample, how would that change how you participate in the project? (By the way, this is just hypothetical – all classifications are included in our sample.)
2. Galaxy Zoo depends on people like you choosing to offer their time to classify galaxies. But, if you were told by someone else, such as a boss or a teacher, to participate in Galaxy Zoo, how would that change how you participate in the project?
3. Galaxy Zoo includes about a million galaxies, most of which have been looked at by only a few people. If Galaxy Zoo instead included galaxies that had been extensively studied before, how would that change how you participate in the project?
4. If Galaxy Zoo did not include a forum, so that you could not interact with other users, how would that change how you participate in the project?
5. Some people on the forum have joked that the site is "addictive" – that they can't help doing "just one more galaxy." If we changed the interface so that you had to request the next galaxy (rather than getting the next one immediately), how would that change how you participate in the project?
6. As you know, we are classifying galaxies that have never been classified before, and you are helping us. But if the site gave you a "right" answer after you finished classifying a galaxy, how would that change how you participate in the project?

Thank you! Now, let me ask for your opinion on some parts of the website. We are interested in your honest assessments – I didn't write the site, so don't worry about hurting my feelings! (4 questions)

**Evaluating the Site**

1. How difficult do you find classifying galaxies? Is there anything that would make it easier?
2. What comments do you have on the Galaxy Zoo interface?
3. From where do you most often use Galaxy Zoo?
4. What suggestions do you have for future versions of the site?

We've reached the end of the interview! Thank you so much for your help.

Here's what happens next. I will now save this transcript, but I will delete your username and refer to you by a code number. Your code number is 47. If you'd like to withdraw from the study at any time, contact the study coordinator, Jordan Raddick (raddick@jhu.edu), and give him your code number. If you have any questions, or you would like a copy of the published research that results from this interview, contact Jordan.

You'll be entered into a prize drawing, as explained in my E-mail to you. We'll contact you by E-mail if you have won (but your E-mail will be kept separate from this transcript).